\pdfoutput=1
\documentclass[%
reprint,
superscriptaddress,
bibnotes,
amsmath,amssymb,
aps,
floatfix,
]{revtex4-1}
\usepackage{graphicx}
\usepackage{dcolumn}
\usepackage{bm}
\usepackage{hyperref}

\usepackage{natbib}
\usepackage{mathrsfs,mathtools,color,wasysym,dsfont,here}
\usepackage{amsthm}
\usepackage{physics}
\usepackage[all]{xy}
\usepackage{tikz}
\usepackage{amsfonts}
\usetikzlibrary{calc}
\usetikzlibrary{arrows}
\usetikzlibrary{decorations.markings,decorations.pathmorphing}

\preprint{}

\newcommand{\C}{\ensuremath{\mathscr{C}}}
\newcommand{\Ce}{\ensuremath{\mathscr{C}_{\text{even}}}}

\newcommand{\Ker}{\ensuremath{\operatorname{Ker}}}

\begin{document}

\title{Rigorous results on the ground state of the attractive SU($N$) Hubbard model}

\author{Hironobu Yoshida}
\email{hironobu-yoshida57@g.ecc.u-tokyo.ac.jp}
\affiliation{Department of Physics, Graduate School of Science, The University of Tokyo, 7-3-1 Hongo, Tokyo 113-0033, Japan}

\author{Hosho Katsura}
\affiliation{Department of Physics, Graduate School of Science, The University of Tokyo, 7-3-1 Hongo, Tokyo 113-0033, Japan}
\affiliation{Institute for Physics of Intelligence, The University of Tokyo, 7-3-1 Hongo, Tokyo 113-0033, Japan}
\affiliation{Trans-scale Quantum Science Institute, The University of Tokyo, 7-3-1, Hongo, Tokyo 113-0033, Japan}
\begin{abstract}
  We study the attractive SU($N$) Hubbard model with particle-hole symmetry. The model is defined on a bipartite lattice with the number of sites $N_A$ $(N_B)$ in the $A$ $(B)$ sublattice. We prove three theorems that allow us to identify the basic ground-state properties: the degeneracy, the fermion number, and the SU($N$) quantum number. We also show that the ground state exhibits charge density wave order when $|N_A-N_B|$ is macroscopically large. The theorems hold for a bipartite lattice in any dimension, even without translation invariance.
\end{abstract}
\maketitle
{\it Introduction.}---
The (fermionic) Hubbard model~\cite{Hubbard1963, Kanamori1963, Gutzwiller1964} is one of the most important models for describing strongly correlated fermions. Despite its apparent simplicity, the model has proved to be notoriously difficult to analyze analytically, and rigorous results are few and far between~\cite{Tasaki1998, essler2005, Lieb2013, Tasaki2020}.\par
Recently, the SU($N$) generalization of the Hubbard model has attracted much attention since it was realized with ultracold atoms in optical lattices~\cite{Abraham1997,Bartenstein2005,Fukuhara2007, Ottenstein2008,Huckans2009,Taie2010,Taie2012,Desalvo2010,Lewenstein2012,Scazza2014,Zhang2014,Cazalilla2014,Pagano2014,Hofrichter2016}. In particular, attractive SU($N$) Hubbard models are predicted to host a variety of exotic phases that do not appear in the SU(2) counterpart, including color superfluid and trion phases with $N=3$~\cite{Zhao2007, PhysRevLett.98.160405, PhysRevB.77.144520, PhysRevLett.98.160405, PhysRevB.77.144520, Titvinidze2011}.\par
In the SU(2) case, the spin-reflection positivity method invented by Lieb~\cite{PhysRevLett.62.1201} is a powerful tool to establish rigorous results. It exploits the symmetry between up-spin and down-spin electrons.
When the interaction is attractive and the number of electrons is even, the ground state was shown to be unique and a spin singlet~\cite{PhysRevLett.62.1201}. When the lattice is bipartite and the difference in the number of sites in the two sublattices is macroscopically large, the coexistence of superconductivity and charge density wave was proved~\cite{Tian1992,PhysRevLett.71.4238,Tian1994,Shen1998}.
This method has also been used to study the ground state of other strongly correlated electron systems~\cite{PhysRevLett.68.1030,PhysRevLett.74.4939,PhysRevB.53.14252,PhysRevB.51.2812,Zhijun_1997,Miyao2012, miyao2020electronphonon,Shen1998,Tsunetugu1997,Yanagisawa1996,Tian2004}, such as the periodic Anderson model and the Kondo lattice model. \par
However, the method in its original form is not applicable to the SU($N$) Hubbard model with $N\geq 3$. Thus, a new approach has to be developed. Here, we use a method based on the Majorana representation of fermions, called Majorana reflection positivity~\cite{Jaffe2015}. While the spin-reflection positivity method uses the symmetry between up-spin and down-spin electrons, the Majorana reflection positivity method relies on the symmetry between two species of Majorana fermions, $\gamma^{(1)}$ and $\gamma^{(2)}$.
It has been used to solve the fermion sign problem in quantum Monte Carlo simulations~\cite{PhysRevB.91.241117,PhysRevLett.115.250601,PhysRevLett.116.250601,Li2019}. For example, the SU($3$) attractive Hubbard model on the honeycomb lattice was numerically studied, and a quantum phase transition from a semimetal to a charge density wave phase was observed~\cite{Xu2019}. The method was also used to discuss the ground state degeneracy of interacting spinless fermions~\cite{Wei2014}.\par
In this letter, we extend the method of Majorana reflection positivity and prove three theorems on the attractive SU($N$) Hubbard model with $N\geq3$. First, we will identify the degeneracy, the fermion number, and the SU($N$) quantum number of the ground state (Theorem 1). This is a natural generalization of Lieb's theorem on the SU($2$) Hubbard model~\cite{PhysRevLett.62.1201}. Next, we will prove an inequality for a correlation function, which is a measure of the charge density wave order (Theorem 2). Finally, combining Theorems 1 and 2, we will show that the system exhibits the charge density wave order when $|N_A-N_B|$ is macroscopically large, where $N_A$ $(N_B)$ is the number of sites in the $A$ $(B)$ sublattice (Theorem 3). This is a natural generalization of Tian's theorem on the SU($2$) Hubbard model~\cite{Tian1994}.

{\it The model and main results.}---
We consider the attractive SU($N$) Hubbard model on a finite bipartite lattice $\Lambda$.  Bipartiteness means that the lattice $\Lambda$ can be divided into two sublattices, $A$ and $B$, and if two sites $x,y\in\Lambda$ belong to the same sublattice, the hopping matrix element $t_{x,y}$ is zero. Let us write the number of sites in the whole lattice $\Lambda$ as $N_{\mathrm{s}}$ and the number of sites in the $A$ $(B)$ sublattice as $N_{A}$ $(N_{B})$. For each site $x \in \Lambda$, we denote by $c^\dagger_{x,\sigma}$ and $c_{x,\sigma}$ the creation and annihilation operators, respectively, of a fermion with flavor $\sigma=1, \cdots, N$. We define the number operators by $n_{x,\sigma} =c^{\dagger}_{x,\sigma} c_{x,\sigma}$ and $n_x=\sum_{\sigma=1}^{N} n_{x,\sigma}$. Let us consider the standard Hamiltonian of the attractive SU($N$) Hubbard model
\begin{align}
  H&=H_\mathrm{hop}+H_\mathrm{int},  \label{eq:ham} \\
  H_\mathrm{hop}&=\sum_{x\in A, y \in B} \sum_{\sigma=1}^{N} t_{x,y} (c^\dagger_{x,\sigma} c_{y,\sigma}+c^\dagger_{y,\sigma} c_{x,\sigma}), \label{eq:hamhop}\\
  H_\mathrm{int}&=\sum_{x\in \Lambda} U_x \left(n_x-\frac{N}{2}\right)^2.
  \label{eq:hamint}
\end{align}
The on-site interactions may depend on sites, as long as $U_x<0$. We assume that the hopping matrix elements $t_{x,y}$ are real. We also assume that the lattice is connected via nonvanishing hopping matrix elements, i.e., for any $x,y\in \Lambda$ such that $x\neq y$, there exists a finite sequence $z_1\cdots z_n \in \Lambda$ with $z_1=x,\,z_n=y$, where $t_{z_j,z_{j+1}}$ are nonvanishing for all $j=1,\cdots n-1$. Note that the Hamiltonian is invariant under the particle-hole transformation $c_{x,\sigma}\to (-1)^{x}c^\dagger_{x,\sigma}$, where $(-1)^x=1$ if $x\in A$ and $(-1)^x=-1$ if $x\in B$. \par
To state our first theorem, let us define SU($N$) singlet states. To this end, we introduce the operators $F^{\sigma, \tau}= \sum_{x\in \Lambda}c^\dagger_{x,\sigma}c_{x,\tau}$. Here, $F^{\sigma, \sigma}$ is the total number operator of fermions with flavor $\sigma$, while $F^{\sigma, \tau}$ ($\sigma \ne \tau$) are flavor-raising and lowering operators. Since all $F^{\sigma,\tau}$ operators commute with the Hamiltonian $H$, it has the global $\mathrm{U}(N)=\mathrm{U}(1)\times \mathrm{SU}(N)$ symmetry. A state $\ket{\Phi_\mathrm{singlet}}$ is an SU($N$) singlet with the total fermion number $N_f$ if $F^{\sigma,\tau}\ket{\Phi_\mathrm{singlet}}=0$ for all $\sigma \neq \tau$ and $F^{\sigma,\sigma}\ket{\Phi_\mathrm{singlet}}=\frac{N_f}{N} \ket{\Phi_\mathrm{singlet}}$ for all $\sigma=1, \cdots, N$~\footnote{These conditions can also be expressed with the SU($N$) version of spin operators. From the operators $F^{\sigma,\tau}$, one can construct new operators as $F=\sum_{\sigma=1}^N F^{\sigma,\sigma}$ and $T^a = \sum_{\sigma,\tau=1}^N \mathcal{T}^a_{\sigma,\tau}F^{\sigma,\tau}$ for $a=1,\cdots N^2-1$, where $\mathcal{T}^a_{\sigma,\tau}$ are the generators of SU$(N)$ Lie algebra. Here, $F$ is the total fermion number operator and $T^a$ are the SU($N$) version of spin operators. The conditions for an SU$(N)$ singlet with fermion number $N_f$ are written as $T^a\ket{\Phi_\mathrm{singlet}}=0$ for all $a=1,\cdots N^2-1$ and $F\ket{\Phi_\mathrm{singlet}}=N_f \ket{\Phi_\mathrm{singlet}}$.}. Our first theorem is stated as follows.
\par
{\it Theorem 1.}---
Consider the attractive SU($N$) Hubbard model with the Hamiltonian \eqref{eq:ham} with $N\geq 3$. When ${N_A\neq N_B}$, there are exactly two ground states in the whole Fock space. The two ground states are SU($N$) singlets and their total fermion numbers are $NN_A$ and $NN_B$, respectively. When ${N_A=N_B}$, there are at most two ground states, each of which is an SU($N$) singlet and whose total fermion number is $NN_A$ $(=NN_B)$.
\par
We can also show an inequality for a correlation function for the ground state. Let us define an operator $S_{x,y}$ for a pair of sites $x,y\in \Lambda$ (including the case $x=y$) as
\begin{equation}
  S_{x,y}=(-1)^{x}(-1)^{y}\left(n_x-\frac{N}{2}\right)\left(n_y-\frac{N}{2}\right),
  \label{eq:def_of_S}
\end{equation}
where $(-1)^x=1$ if $x\in A$ and $(-1)^x=-1$ if $x\in B$.
Then, our second theorem is stated as follows.
\par
{\it Theorem 2.}---
Under the same conditions as in Theorem 1, we have for any ground state $\ket{\Phi_\mathrm{GS}}$ and for $x,y\in \Lambda$ that
\begin{equation}
  \bra{\Phi_\mathrm{GS}}S_{x,y}\ket{\Phi_\mathrm{GS}}>0.
  \label{eq:main2}
\end{equation}
\par
The correlation function $\bra{\Phi_\mathrm{GS}}S_{x,y}\ket{\Phi_\mathrm{GS}}$ is a measure of the charge density wave order. Note that this inequality does not necessarily imply the presence of the long-range order in the thermodynamic limit.
\par
However, when $|N_A-N_B|$ is macroscopically large, we can prove the presence of the long-range order. Assume that $|N_A-N_B|=a N_{\mathrm{s}}$ with a constant $a$ such that ${0\leq a<1}$. Note that $N_{\mathrm{s}}=N_A+N_B$. The order parameter for the charge density wave is
\begin{equation}
  S_\mathrm{CDW}=\sum_{x\in \Lambda} (-1)^x\left(n_x-\frac{N}{2}\right).
\end{equation}
Then, our third theorem is stated as follows.
\par
{\it Theorem 3.}---
Under the same conditions as in Theorem 1, we have for any ground state $\ket{\Phi_\mathrm{GS}}$ that
\begin{equation}
  \bra{\Phi_\mathrm{GS}}(S_\mathrm{CDW})^2\ket{\Phi_\mathrm{GS}}> \left(\frac{aNN_\mathrm{s}}{2}\right)^2.
  \label{eq:main3}
\end{equation}
Since the right hand side of the inequality \eqref{eq:main3} is proportional to $N_\mathrm{s}^2$  for $0< a<1$, this theorem shows that the system has long-range order.
\par
Theorem 3 follows from Theorems 1 and 2.
\par
{\it Proof of Theorem 3.}---
First one finds
\begin{equation}
  (S_\mathrm{CDW})^2 =\sum_{x,y\in \Lambda} S_{x,y}.
  \label{eq:proof3-1}
\end{equation}
By using the inequality \eqref{eq:main2},
\begin{equation}
  \begin{split}
    &\bra{\Phi_\mathrm{GS}}\sum_{x,y\in \Lambda} S_{x,y}\ket{\Phi_\mathrm{GS}} \\
    &>\bra{\Phi_\mathrm{GS}}\sum_{x,y\in \Lambda} (-1)^x(-1)^yS_{x,y} \ket{\Phi_\mathrm{GS}} \\
    &=\bra{\Phi_\mathrm{GS}}\left[\sum_{x,y\in \Lambda}\left(n_x-\frac{N}{2}\right)\left(n_y-\frac{N}{2}\right)\right]\ket{\Phi_\mathrm{GS}} \\
    &=\bra{\Phi_\mathrm{GS}}\left[\sum_{x\in \Lambda}\left(n_x-\frac{N}{2}\right)\right]^2\ket{\Phi_\mathrm{GS}}.
    \label{eq:proof3-2}
  \end{split}
\end{equation}
From Theorem 1, the total fermion number of the ground state is $NN_A$ or $NN_B$. Substituting $\sum_{x\in \Lambda} n_x=NN_A$ or $NN_B$ into \eqref{eq:proof3-2} and using $|N_A-N_B|=aN_{\mathrm{s}}$, we obtain \eqref{eq:main3}.
\hspace{\fill}$\blacksquare$
To prove Theorems 1 and 2, we use a matrix representation of eigenstates introduced by Wei et al.~\cite{Wei2014}. First, we will show the following lemma.
\par
{\it Lemma 4.}---
Consider the attractive SU($N$) Hubbard model with the Hamiltonian \eqref{eq:ham} with $N\geq 3$. When $NN_{\mathrm{s}}$ is odd, there are exactly two ground states. When $NN_{\mathrm{s}}$ is even, there are at most two ground states.
\par
In the following discussion, we only consider the case where $NN_{\mathrm{s}}$ is odd. For even $NN_{\mathrm{s}}$, see the Supplemental Material~\footnote{See Supplemental Material for detailed derivations and proofs}.
\par
{\it The Majorana representation.}---
A complex fermion can be decomposed into two Majorana fermions. We define $\gamma^{(1)}_{x,\sigma}=c_{x,\sigma}+c^\dagger_{x,\sigma}$, $\gamma^{(2)}_{x,\sigma}=-i(c_{x,\sigma}-c^\dagger_{x,\sigma})$ at sublattice $A$ and $\gamma^{(1)}_{x,\sigma}=-i(c_{x,\sigma}-c^\dagger_{x,\sigma})$, $\gamma^{(2)}_{x,\sigma}=c_{x,\sigma}+c^\dagger_{x,\sigma}$ at sublattice $B$. They satisfy the relations
\begin{equation}
  \gamma^{(j)\dagger}_{x,\sigma} = \gamma^{(j)}_{x,\sigma},\quad \left\{\gamma^{(j)}_{x,\sigma},\gamma^{(k)}_{y,\tau} \right\}=2\delta_{j,k}\delta_{x,y}\delta_{\sigma, \tau}
\end{equation}
for all $x,y\in \Lambda$, $\sigma,\tau=1,\cdots, N$, $j,k=1,2$. Using the Majorana representation, we can rewrite \eqref{eq:hamhop} and \eqref{eq:hamint} as
\begin{equation}
  H_\mathrm{hop}=\sum_{x\in A, y\in B} \sum_{\sigma=1}^{N} t_{x,y} \left(\frac{i}{2}\gamma^{(1)}_{x,\sigma} \gamma^{(1)}_{y,\sigma}-\frac{i}{2}\gamma^{(2)}_{x,\sigma} \gamma^{(2)}_{y,\sigma}\right),\label{eq:Majoranahop}
\end{equation}
\begin{equation}
  H_\mathrm{int}=\sum_{x\in \Lambda} \sum_{\sigma,\tau=1}^{N} U_x \left(\frac{i}{2}\gamma^{(1)}_{x,\sigma} \gamma^{(1)}_{x,\tau}\right)\left(-\frac{i}{2}\gamma^{(2)}_{x,\sigma} \gamma^{(2)}_{x,\tau}\right).
  \label{eq:Majoranaint}
\end{equation}
\par
The operators on the whole Fock space form a complex vector space. We write this vector space as $\mathscr{O}$. Note that the dimension of $\mathscr{O}$ is $2^{N N_{\mathrm{s}}}$. We introduce the Hilbert-Schmidt inner product for $O_1,O_2\in \mathscr{O}$ as
\begin{equation}
  (O_1, O_2)= \frac{1}{2^{N N_{\mathrm{s}}}} \Tr\left[O_1^\dagger O_2\right].
  \label{eq:def_of_ip}
\end{equation}
Then, operators defined by
\begin{align}
  \Gamma_\alpha^{(1)}&= i^{\lfloor l(\alpha)/2\rfloor}\gamma_{x_1,\sigma_1}^{(1)}\cdots \gamma_{x_{l(\alpha)},\sigma_{l(\alpha)}}^{(1)},\label{eq:def_of_Majorana_basis1} \\
  \Gamma_\alpha^{(2)}&= (-i)^{\lfloor l(\alpha)/2\rfloor}\gamma_{x_1,\sigma_1}^{(2)}\cdots \gamma_{x_{l(\alpha)},\sigma_{l(\alpha)}}^{(2)}.
  \label{eq:def_of_Majorana_basis2}
\end{align}
form an orthonormal basis of $\mathscr{O}$. Here, $\alpha= \bigl((x_1,\sigma_1),\cdots,(x_{l(\alpha)},\sigma_{l(\alpha)})\bigr)$ denotes a subset of $\Lambda \times \{1,2,\cdots N\}$ ordered according to an arbitrary order introduced in $\Lambda \times \{1,2,\cdots N\}$. We wrote the length of $\alpha$ as $l(\alpha)$, and $\lfloor l(\alpha)/2\rfloor$ is the largest integer less than or equal to $l(\alpha)/2$.
We write the set of $\alpha$ as $\C$, and the set of even (odd)-length $\alpha$ as $\C_{\text{even(odd)}}$. Here, $|\C|=2^{NN_{\mathrm{s}}}$ and $|\C_{\text{even}}|=|\C_{\text{odd}}|=2^{NN_{\mathrm{s}}-1}$~\footnote{$|S|$ denotes the number of elements in $S$}. We also define the parity operators,
\begin{align}
  \Delta^{(1)}&= i^{\lfloor NN_{\mathrm{s}}/2\rfloor}\prod_{x\in \Lambda}\prod_{\sigma=1}^{N}\gamma_{x,\sigma}^{(1)} , \\
  \Delta^{(2)}&= (-i)^{\lfloor NN_{\mathrm{s}}/2\rfloor}\prod_{x\in \Lambda}\prod_{\sigma=1}^{N}\gamma_{x,\sigma}^{(2)},
\end{align}
 which commute with the Hamiltonian. Here, we assumed that the product is ordered in the same order as $\Gamma_\alpha^{(1)}$ and $\Gamma_\alpha^{(2)}$. Note that $\Delta^{(1)}$ commutes (anticommutes) with $\Delta^{(2)}$ when $NN_{\mathrm{s}}$ is even (odd), and $(\Delta^{(1)})^2=(\Delta^{(2)})^2=1$.
\par
Next, define {\it eigenoperators} of the Hamiltonian.
\par
{\it Definition 5.}---
An operator $O \in \mathscr{O}$ is said to be an eigenoperator of $H$ with eigenvalue $E$ when $HO=OH=EO$. We denote by $\mathscr{O}^E$ the subspace of $\mathscr{O}$ spanned by the eigenoperators of $H$ with eigenvalue $E$.
\par
Let us consider the relation between the eigenoperator formalism and the ordinary eigenvector formalism. Let $\bigl\{\ket{E,j}|j=1,\cdots n_E\bigr\}$ be the complete set of eigenvectors of $H$ with eigenvalue $E$. Then, the subspace of $\mathscr{O}$ spanned by $\bigl\{\ket{E,j}\bra{E,k}\big|j,k=1,\cdots n_E\bigr\}$ corresponds to $\mathscr{O}^E$. Therefore, if the degeneracy of the ground state eigenvectors is $n_E$, the degeneracy of the ground state eigenoperators is $n_E^2$.
\par
The eigenoperator can be decomposed into four sectors, because the Hamiltonian preserves the parity (even or odd) of the number of $\gamma^{(1)}$ and $\gamma^{(2)}$, respectively.
\begin{equation}
  \mathscr{O}=\mathscr{O}_\text{even,even}\oplus\mathscr{O}_\text{even,odd}\oplus\mathscr{O}_\text{odd,even}\oplus\mathscr{O}_\text{odd,odd},
\end{equation}
where $\mathscr{O}_{\text{even (odd)},\text{even (odd)}}$ is the subspace of $\mathscr{O}$ spanned by  $\left\{\Gamma^{(1)}_{\alpha}\Gamma^{(2)}_{\beta} \big|\alpha\in \C_\text{even(odd)},\,\beta\in \C_\text{even(odd)}\right\}$.
\par
When $NN_{\mathrm{s}}$ is odd, each parity operator $\Delta^{(1)}$ and $\Delta^{(2)}$ contains an odd number of Majorana operators. Thus they define maps between different sectors. For example, if $O$ is in the even-even sector, $\Delta^{(1)}O$, $\Delta^{(2)}O$, $\Delta^{(1)}\Delta^{(2)}O$ are in the odd-even sector, even-odd sector, odd-odd sector, respectively. Furthermore, these are maps between eigenoperators with the same energy because they commute with the Hamiltonian. Therefore, if the ground state eigenoperator is unique in the even-even sector, the total degeneracy of the ground state eigenoperators is four. This means that the ground state eigenvectors are two-fold degenerate.
\par
In the following discussion, we focus on the even-even sector. In this sector, an operator is expressed as
\begin{equation}
  O(W)=\sum_{\alpha,\beta \in \C_\text{even}} W_{\alpha,\beta} \Gamma_\alpha^{(1)} \Gamma_\beta^{(2)},
  \label{eq:expansion}
\end{equation}
where $W$ is a $|\Ce|\times |\Ce|$ matrix. This matrix representation plays an essential role in the proof.
\par
Let an operator ${O}(W)\in \mathscr{O}_\text{even,even}$ be an eigenoperator of ${H}$ with eigenvalue $E$.  Then $W$ satisfies the following two equations. The derivation is summarized in the Supplemental Material~\cite{Note2}.
\begin{equation}
  KW+WK+\sum_{x\in \Lambda} \sum_{\sigma,\tau=1}^{N} U_x L_{xx,\sigma \tau} W L_{xx,\sigma \tau} = EW,
  \label{eq:eigenequation}
\end{equation}
\begin{equation}
  K^{\top}W+WK^{\top}+\sum_{x\in \Lambda}\sum_{\sigma,\tau=1}^{N} U_x L_{xx,\sigma \tau}^{\top} W L_{xx,\sigma \tau}^{\top} = EW,
  \label{eq:transeigenequation}
\end{equation}
where $\top$ denotes the transpose. $L_{xy,\sigma\tau}$ and $K$ are $|\Ce|\times |\Ce|$ Hermitian matrices defined by
\begin{gather}
  (L_{xy,\sigma\tau})_{\alpha,\beta}=\left(\Gamma^{(1)}_\alpha, \frac{i}{2}\gamma_{x,\sigma}^{(1)}\gamma_{y,\tau}^{(1)}\Gamma^{(1)}_\beta\right), \label{eq:defL}\\
  (K)_{\alpha,\beta}= \sum_{x\in A, y \in B} \sum_{\sigma=1}^{N}  t_{x,y} (L_{xy,\sigma\sigma})_{\alpha,\beta}\label{eq:defK}.
\end{gather}
Since $L_{xy,\sigma\tau}$ and $K$ are Hermitian, we find $W^\dagger$ also satisfies \eqref{eq:eigenequation} and \eqref{eq:transeigenequation} and hence ${O}(W^{\dagger})$ is also an eigenoperator with eigenvalue $E$. Thus $W$ can be symmetrized or antisymmetrized to be Hermitian.
\par
Let us define the normalization condition for operators $O\in \mathscr{O}$ as $(O,O)=1$, where the inner product is defined by \eqref{eq:def_of_ip}. Since $(O(W),O(W))=\Tr[W^\dagger W]$, the normalization condition for Hermitian $W$ is $\Tr[W^2]=1$. Under the normalization condition, the expectation value of $H$ with respect to $O$ is defined by $\left(O,H O\right)$. We define $E(W) = \left(O(W),H O(W)\right)$ for a normalized Hermitian matrix $W$. Then $E(W)$ is calculated as
\begin{equation}
  E(W)= 2 \text{Tr}[KW^2]+\sum_{x\in \Lambda} \sum_{\sigma,\tau=1}^{N} U_x \text{Tr}[W L_{xx,\sigma \tau} W L_{xx,\sigma \tau}].
\end{equation}
For a Hermitian matrix $W$ , by diagonalizing it by a unitary matrix $U$ like $W = U D U^\dagger,\,D=\operatorname{diag}(\lambda_1,\dots \lambda_{|\C_{\text{even}}|})$, we can define a new matrix $|W|= U |D| U^\dagger$, where $|D|=\operatorname{diag}(|\lambda_1|,\dots |\lambda_{|\C_{\text{even}}|}|)$. Then we have $E(|W|)\leq E(W).$ If $W$ is normalized, $|W|$ is also normalized. Therefore, by the variational principle, if $O(W)$ is a ground state eigenoperator, then $O(|W|)$ is also a ground state eigenoperator.
\par
{\it Implication of connectivity.}---
Here we prove the following lemmas, which is essential in the proof of Lemma 4 and Theorem 2.
\par
{\it Lemma 6. }---
Consider the attractive SU($N$) Hubbard model with $N\geq 3$. If a positive semidefinite matrix $W$ satisfies \eqref{eq:eigenequation} and \eqref{eq:transeigenequation}, then $W$ is either positive definite or zero.
\par
See the Supplemental Material for a proof~\cite{Note2}. The condition $N\geq 3$ comes from this lemma. From Lemma 6, we can show the following lemma.
\par
{\it Lemma 7. }---
Consider the attractive SU($N$) Hubbard model with $N\geq 3$. If a ground state eigenoperator is $O(W)$, then $W$ is either positive or negative definite.
\par
{\it Proof of Lemma 7.}---
Let $O(W) \in \mathscr{O}_\text{even,even}$ be a ground state eigenoperator in the even-even sector. Then $O(|W|)$ is also a ground state eigenoperator. Thus $|W|-W$ is a positive semidefinite matrix which satisfies \eqref{eq:eigenequation} and \eqref{eq:transeigenequation} with $E=E_\mathrm{GS}$. Here, $E_\mathrm{GS}$ is the ground state energy. From Lemma 6, $|W|-W$ is either positive definite or zero. If $|W|-W$ is positive definite, all eigenvalues of $W$ is strictly negative, which means that $W$ is negative definite. If $|W|-W$ is zero, then $W=|W|$. By using Lemma 6, $|W|$ is positive definite because $|W|$ is a nonvanishing positive semidefinite matrix which satisfies \eqref{eq:eigenequation} and \eqref{eq:transeigenequation}. Thus $W$ is also positive definite. Therefore, $W$ is positive or negative definite.\hspace{\fill}$\blacksquare$ \par
We can prove Lemma 4 from Lemma 7.\par
{\it Proof of Lemma 4 for odd $NN_{\mathrm{s}}$.}---
Suppose that the ground state eigenoperators in the even-even sector are degenerate. Then, we pick two orthogonal ground state eigenoperators $O(W_1)$ and $O(W_2)$. Then, $\Tr[W_1^\dagger W_2]=(O(W_1),O(W_2))=0$. But Lemma 7 implies that $\Tr[W_1^\dagger W_2]\neq0$~\footnote{See page 364 of Ref.~\cite{Tasaki2020} for a proof.}. Since this is a contradiction, the ground state eigenoperator is unique in the even-even sector. Therefore, there are exactly two ground states in total.\hspace{\fill}$\blacksquare$
\par
We will complete the proof of Theorem 1 for odd $NN_\mathrm{s}$ by identifying the SU($N$) quantum number and the total fermion number of the ground states.
\par
{\it Proof of Theorem 1 for odd $NN_{\mathrm{s}}$.}---
First, we determine the SU($N$) quantum numbers of the ground states. Note that the ground state degeneracy in an SU($N$) invariant model is at least $N$ unless the ground states are SU($N$) singlets. This, together with Lemma 4, implies that the two ground states are SU($N$) singlets.
\par
To determine the fermion number, we consider a toy model on the same lattice with long-range interactions. The Hamiltonian of the model is
\begin{equation}
  H_{\mathrm{toy}} = \sum_{x\in A, y\in B}\left(n_x-\frac{N}{2}\right)\left(n_y-\frac{N}{2}\right).
  \label{eq:toyHam}
\end{equation}
The ground states of the model are two-fold degenerate. Let us write the two ground states as $\ket{\Phi_{\pm}}$. Then,
\begin{equation}
  n_x \ket{\Phi_{\pm}} = \left\{
  \begin{array}{cl}
    \displaystyle \frac{(N\pm N)}{2} \ket{\Phi_{\pm}} &
    \mathrm{if}\ x \in A, \\
    & \\
    \displaystyle \frac{(N\mp N)}{2} \ket{\Phi_{\pm}} &
    \mathrm{if}\ x \in B.
  \end{array}
  \right.
  \label{eq:gs_of_toyHam}
\end{equation}
As shown in Ref.~\cite{Zhao2007} (see the Supplemental Material~\cite{Note2}), $\ket{\Phi_{+}}$ and $\ket{\Phi_{-}}$ are also the ground states of the attractive SU($N$) Hubbard model in the large-$U_x$ limit. The fermion numbers of $\ket{\Phi_{+}}$ and $\ket{\Phi_{-}}$ are $NN_A$ and $NN_B$, respectively. The eigenoperators are written as $O_\pm = \ket{\Phi_{\pm}}\bra{\Phi_{\pm}} = 2^{-NN_\mathrm{s}}\prod_{x\in \Lambda}\prod_{\sigma=1}^N \left(1\pm i\gamma^{(1)}_{x,\sigma}\gamma^{(2)}_{x,\sigma}\right)$.
Then, ${2^{NN_\mathrm{s}-1}(O_++O_-)}$ is in $\mathscr{O}_\mathrm{even,even}$ sector and written as $O(I)$, where $I$ is the identity matrix of size $|\Ce|$. Let $O(W_\mathrm{GS}) \in \mathscr{O}_\mathrm{even,even}$ be the ground state eigenoperator of the original Hamiltonian $H$. By Lemma 7,
$\left(O(I),O\left(W_\mathrm{GS}\right)\right)=\Tr\left[{W_\mathrm{GS}}\right]\neq0$, because $W_\mathrm{GS}$ is positive or negative definite. Suppose we expand $O\left(W_\mathrm{GS}\right)$ in an orthonormal basis of $\mathscr{O}$ including $\ket{\Phi_{+}}\bra{\Phi_{+}}$ and $\ket{\Phi_{-}}\bra{\Phi_{-}}$. Since $\left(O(I),O\left(W_\mathrm{GS}\right)\right)\neq0$, the coefficient of either $\ket{\Phi_{+}}\bra{\Phi_{+}}$ or $\ket{\Phi_{-}}\bra{\Phi_{-}}$ is nonzero.
\par
Let $P_{A (B)}$ be the projection operator onto states with $NN_A$ $(NN_B)$ fermions. Then, either $P_{A}O\left(W_\mathrm{GS}\right)P_{A}$ or $P_{B}O\left(W_\mathrm{GS}\right)P_{B}$ is nonzero. Since $P_{A (B)}$ commutes with the Hamiltonian, the projected operators are also ground state eigenoperators. Therefore, there is a ground state whose fermion number is $NN_A$ or $NN_B$. Due to the particle-hole symmetry, if there is a ground state with the fermion number $NN_A$ ($NN_B$), there must be another ground state with the fermion number $NN_B$ ($NN_A$). Note that $NN_A\neq NN_B$ when $NN_\mathrm{s}$ is odd. This, together with Lemma 4, implies that there are exactly two ground states, and the fermion numbers of the two ground states are $NN_A$ and $NN_B$, respectively.
\hspace{\fill}
$\blacksquare$
\par
To prove Theorem 2, we use the following lemma~\footnote{Here, we provide a proof of Lemma 8. Take the orthonormal basis $\{\bm{u}_j\}_{j=1,\cdots D}$, which consists of the eigenvectors of $M$, i.e., $M\bm{u}_j=\lambda_j\bm{u}_j$. If $M$ is positive (negative) definite, $\lambda_j>0$ ($\lambda_j<0$) for all $j=1,\cdots D$. Then ${\Tr [MM'MM']} = {\sum_{j,k=1}^{D} \bm{u}_j^\dagger MM'M\bm{u}_k\bm{u}_k^\dagger M'\bm{u}_j} = {\sum_{j,k=1}^{D} \lambda_j \lambda_k \left|\bm{u}_k^\dagger M'\bm{u}_j\right|^2} >0,$
because $\left|\bm{u}_k^\dagger M'\bm{u}_j\right|^2>0$  for some $j$ and $k$.
\hspace{\fill}$\blacksquare$}.
\par
{\it Lemma 8.}---
Let $M,M'$ be $D\times D$ Hermitian matrices. If $M$ is positive or negative definite and $M'$ is nonvanishing, then
\begin{equation}
  \Tr [MM'MM']>0.
\end{equation}
\par
{\it Proof of Theorem 2 for odd $NN_{\mathrm{s}}$.}---
We consider the ground state expectation value of the operator $S_{x,y}$ defined as \eqref{eq:def_of_S}. First, using the Majorana representation, $S_{x,y}$ is expressed as
\begin{equation}
  S_{x,y}=\sum_{\sigma,\tau=1}^N \left(\frac{i}{2}\gamma^{(1)}_{x,\sigma}\gamma^{(1)}_{y,\tau}\right)\left(-\frac{i}{2}\gamma^{(2)}_{x,\sigma}\gamma^{(2)}_{y,\tau}\right).
\end{equation}
\par
When $NN_{\mathrm{s}}$ is odd, the ground state eigenoperators are four-fold degenerate in total and in $\mathscr{O}_\mathrm{even,even}$, $\mathscr{O}_\mathrm{even,odd}$, $\mathscr{O}_\mathrm{odd,even}$, $\mathscr{O}_\mathrm{odd,odd}$ sectors, respectively. Let us first consider $\mathscr{O}_\mathrm{even,even}$ sector. Assume that the ground state eigenoperator in this sector is expressed as $O(W_\mathrm{GS})$, where $W_\mathrm{GS}$ is a Hermitian matrix.
Then, the expectation value for $S_{x,y}$ is calculated as
\begin{equation}
  (O(W_\mathrm{GS}),S_{x,y}O(W_\mathrm{GS}))= \sum_{\sigma,\tau=1}^N\Tr\left[W_\mathrm{GS} L_{xy,\sigma\tau} W_\mathrm{GS}L_{xy,\sigma\tau}\right] ,
\end{equation}
where $L_{xy,\sigma\tau}$ is a Hermitian matrix defined as \eqref{eq:defL}. From Lemma 7, $W_\mathrm{GS}$ is positive or negative definite. Using Lemma 8, we obtain $\Tr\left[W_\mathrm{GS}L_{xy,\sigma\tau} W_\mathrm{GS}L_{xy,\sigma\tau}\right]>0$. Thus one finds  $\left(O(W_\mathrm{GS}),S_{x,y}O(W_\mathrm{GS})\right)>0$.
\par
We next note that the ground state eigenoperators in $\mathscr{O}_\mathrm{even,odd}$, $\mathscr{O}_\mathrm{odd,even}$, $\mathscr{O}_\mathrm{odd,odd}$ sectors are $\Delta^{(1)}O(W_\mathrm{GS})$, $\Delta^{(2)}O(W_\mathrm{GS})$, $\Delta^{(1)}\Delta^{(2)}O(W_\mathrm{GS})$, respectively. Since both of $\Delta^{(1)}$ and $\Delta^{(2)}$ commute with $S_{x,y}$, the expectation value of $S_{x,y}$ does not depend on the choice of the ground state.
If two operators $O_1,O_2 \in \mathscr{O}$ are in different sectors, $(O_1, S_{x,y}O_2)$ is zero because each term of $S_{x,y}$ has the even number of $\gamma^{(1)}$ and $\gamma^{(2)}$ fermions. Therefore, we obtain \eqref{eq:main2} for any ground state when $NN_{\mathrm{s}}$ is odd.
\hspace{\fill}$\blacksquare$
\par
{\it Summary.}---
We presented the degeneracy, the fermion number, and the SU($N$) quantum number of the ground state of the attractive SU($N$) Hubbard model with particle-hole symmetry. We also showed that the ground state has the charge density wave long-range order when $|N_A-N_B|$ is macroscopically large. One can easily extend our results to include attractive (repulsive) interactions between two sites in the same (different) sublattice. Although we focused on a model with SU($N$) symmetry, we expect that our approach will find further applications to $N$-component fermionic models with flavor-dependent hopping and interaction~\cite{Azaria2009}. It would also be interesting to consider the application of the method to other multi-component fermionic systems such as SO(5) symmetric models~\cite{doi:10.1142/S0217984906012213}.
\par
\medskip
H. K. was supported in part by JSPS Grant-in-Aid for Scientific Research on Innovative Areas: No. JP20H04630, JSPS KAKENHI Grant No. JP18K03445, and the Inamori Foundation. H. Y. acknowledges the support of the Forefront Physics and Mathematics Program to Drive Transformation (FoPM).
\par
\bibliographystyle{apsrev4-1}
\bibliography{reference}
\widetext
\renewcommand{\theequation}{S\arabic{equation}}
\setcounter{equation}{0}

\begin{center}
	\textbf{\large Supplemental Material for: ``Rigorous results on the ground state of the attractive SU($N$) Hubbard model''}
\end{center}

\section{The derivation of Equations (20) and (21)}

Let $O(W)=\sum_{\alpha,\beta \in \C_\text{even}} W_{\alpha,\beta} \Gamma_\alpha^{(1)} \Gamma_\beta^{(2)}$ be an eigenoperator of $H$ with eigenvalue $E$. By definition, $O(W)$ satisfy $HO(W)=O(W)H=EO$. We will rewrite it by using the Majorana representation of the Hamiltonian (11), (12). First, since $\left(\Gamma_{\alpha}^{(1)} \Gamma_{\beta}^{(2)},\Gamma_{\alpha'}^{(1)} \Gamma_{\beta'}^{(2)}\right)=\delta_{\alpha,\alpha'}\delta_{\beta,\beta'}$, one finds
\begin{equation}
  \begin{split}
    \frac{i}{2}\gamma^{(1)}_{x,\sigma}\gamma^{(1)}_{y,\sigma} \sum_{\alpha,\beta \in \C_\text{even}} W_{\alpha,\beta} \Gamma_\alpha^{(1)} \Gamma_\beta^{(2)}
    &= \sum_{\alpha,\beta,\alpha',\beta' \in \C_\text{even}} W_{\alpha,\beta}\left(\Gamma_{\alpha'}^{(1)} \Gamma_{\beta'}^{(2)},\,\frac{i}{2}\gamma^{(1)}_{x,\sigma}\gamma^{(1)}_{y,\sigma}\Gamma_\alpha^{(1)} \Gamma_\beta^{(2)}\right) \Gamma_{\alpha'}^{(1)} \Gamma_{\beta'}^{(2)} \\
    &= \sum_{\alpha,\beta,\alpha',\beta' \in \C_\text{even}}  W_{\alpha,\beta}\delta_{\beta,\beta'}\left(\Gamma_{\alpha'}^{(1)}  ,\,\frac{i}{2}\gamma^{(1)}_{x,\sigma}\gamma^{(1)}_{y,\sigma}\Gamma_\alpha^{(1)} \right) \Gamma_{\alpha'}^{(1)} \Gamma_{\beta'}^{(2)} \\
    &= \sum_{\alpha,\beta,\alpha' \in \C_\text{even}}  \left(\Gamma_{\alpha}^{(1)}  ,\,\frac{i}{2}\gamma^{(1)}_{x,\sigma}\gamma^{(1)}_{y,\sigma}\Gamma_{\alpha'}^{(1)} \right) W_{\alpha',\beta} \Gamma_{\alpha}^{(1)} \Gamma_{\beta}^{(2)} \\
    &= \sum_{\alpha,\beta,\alpha' \in \C_\text{even}}  (L_{xy,\sigma\sigma})_{\alpha,\alpha'} W_{\alpha',\beta} \Gamma_{\alpha}^{(1)} \Gamma_{\beta}^{(2)},
  \end{split}
  \label{eq:rep1}
\end{equation}
where $L$ is defined as
\begin{equation}
  (L_{xy,\sigma\tau})_{\alpha,\beta}=\left(\Gamma^{(1)}_\alpha, \frac{i}{2}\gamma_{x,\sigma}^{(1)}\gamma_{y,\tau}^{(1)}\Gamma^{(1)}_\beta\right).
\end{equation}
If $(x,\sigma)\neq(y,\tau)$, then $L_{xy,\sigma\tau}$ is Hermitian because
\begin{equation}
  \begin{split}
    \Tr\left[\Gamma^{(1)}_\alpha\frac{i}{2}\gamma_{x,\sigma}^{(1)}\gamma_{y,\tau}^{(1)}\Gamma^{(1)}_\beta\right]
    =\Tr\left[\left(\Gamma^{(1)}_\beta\frac{(-i)}{2}\gamma_{y,\tau}^{(1)}\gamma_{x,\sigma}^{(1)}\Gamma^{(1)}_\alpha\right)^\dagger\right]
    =\Tr\left[\left(\Gamma^{(1)}_\beta\frac{i}{2}\gamma_{x,\sigma}^{(1)}\gamma_{y,\tau}^{(1)}\Gamma^{(1)}_\alpha\right)^\dagger\right]
    = \Tr\left[\Gamma^{(1)}_\beta\frac{i}{2}\gamma_{x,\sigma}^{(1)}\gamma_{y,\tau}^{(1)}\Gamma^{(1)}_\alpha\right]^* .
  \end{split}
\end{equation}
Similarly,
\begin{equation}
  \begin{split}
    -\frac{i}{2}\gamma^{(2)}_{x,\sigma}\gamma^{(2)}_{y,\sigma} \sum_{\alpha,\beta \in \C_\text{even}} W_{\alpha,\beta} \Gamma_\alpha^{(1)} \Gamma_\beta^{(2)}
    &= -\sum_{\alpha,\beta,\alpha',\beta' \in \C_\text{even}} W_{\alpha,\beta}\left(\Gamma_{\alpha'}^{(1)} \Gamma_{\beta'}^{(2)},\,\frac{i}{2}\gamma^{(2)}_{x,\sigma}\gamma^{(2)}_{y,\sigma}\Gamma_\alpha^{(1)} \Gamma_\beta^{(2)}\right) \Gamma_{\alpha'}^{(1)} \Gamma_{\beta'}^{(2)} \\
    &= -\sum_{\alpha,\beta,\alpha',\beta' \in \C_\text{even}}  W_{\alpha,\beta}\delta_{\alpha,\alpha'}\left(\Gamma_{\beta'}^{(2)}  ,\,\frac{i}{2}\gamma^{(2)}_{x,\sigma}\gamma^{(2)}_{y,\sigma}\Gamma_\beta^{(2)} \right) \Gamma_{\alpha'}^{(1)} \Gamma_{\beta'}^{(2)} \\
    &= -\sum_{\alpha,\beta,\beta' \in \C_\text{even}}  \left(\Gamma_{\beta}^{(2)}  ,\,\frac{i}{2}\gamma^{(2)}_{x,\sigma}\gamma^{(2)}_{y,\sigma}\Gamma_{\beta'}^{(2)} \right) W_{\alpha,\beta'} \Gamma_{\alpha}^{(1)} \Gamma_{\beta}^{(2)}. \\
    \label{eq:twotransform1}
  \end{split}
\end{equation}
The inner product in the final line of \eqref{eq:twotransform1} can be rewritten as
\begin{equation}
  \left(\Gamma_{\beta}^{(2)}  ,\,\frac{i}{2}\gamma^{(2)}_{x,\sigma}\gamma^{(2)}_{y,\sigma}\Gamma_{\beta'}^{(2)} \right)
  = (-1)^\frac{-l(\beta)+l(\beta')}{2}  \left(\Gamma_{\beta}^{(1)}  ,\,\frac{i}{2}\gamma^{(1)}_{x,\sigma}\gamma^{(1)}_{y,\sigma}\Gamma_{\beta'}^{(1)} \right) ,
  \label{eq:twotransform2}
\end{equation}
where we used (14), (15), $\beta,\beta'\in \C_{\text{even}}$ and the symmetry between $\gamma^{(1)}$ and $\gamma^{(2)}$ fermions. Since $\left(\Gamma_{\beta'}^{(1)}  ,\,\frac{i}{2}\gamma^{(1)}_{x,\sigma}\gamma^{(1)}_{y,\sigma}\Gamma_{\beta}^{(1)} \right)$ is real (pure imaginary) for odd (even) $\frac{l(\beta)+l(\beta')}{2}$, we get
\begin{equation}
  \begin{split}
    \left(\Gamma_{\beta}^{(1)}  ,\,\frac{i}{2}\gamma^{(1)}_{x,\sigma}\gamma^{(1)}_{y,\sigma}\Gamma_{\beta'}^{(1)} \right)
    &=  \left(\Gamma_{\beta'}^{(1)}  ,\,\frac{i}{2}\gamma^{(1)}_{x,\sigma}\gamma^{(1)}_{y,\sigma}\Gamma_{\beta}^{(1)} \right)^* \\
    &=  -(-1)^{\frac{l(\beta)+l(\beta')}{2}}\left(\Gamma_{\beta'}^{(1)}  ,\,\frac{i}{2}\gamma^{(1)}_{x,\sigma}\gamma^{(1)}_{y,\sigma}\Gamma_{\beta}^{(1)} \right) .
    \label{eq:twotransform3}
  \end{split}
\end{equation}
Combining \eqref{eq:twotransform1}, \eqref{eq:twotransform2} and \eqref{eq:twotransform3}, one finds
\begin{equation}
  -\frac{i}{2}\gamma^{(2)}_{x,\sigma}\gamma^{(2)}_{y,\sigma} \sum_{\alpha,\beta \in \C_\text{even}} W_{\alpha,\beta} \Gamma_\alpha^{(1)} \Gamma_\beta^{(2)}=\sum_{\alpha,\beta,\beta' \in \C_\text{even}}   W_{\alpha,\beta'}(L_{xy,\sigma\sigma})_{\beta',\beta} \Gamma_{\alpha}^{(1)} \Gamma_{\beta}^{(2)}.
  \label{eq:rep2}
\end{equation}
The interaction terms can be rewritten as
\begin{equation}
  \begin{split}
    &\left(\frac{i}{2}\gamma^{(1)}_{x,\sigma} \gamma^{(1)}_{x,\tau}\right)\left(-\frac{i}{2}\gamma^{(2)}_{x,\sigma} \gamma^{(2)}_{x,\tau}\right) \sum_{\alpha,\beta \in \C_\text{even}} W_{\alpha,\beta} \Gamma_\alpha^{(1)} \Gamma_\beta^{(2)} \\
    &= \sum_{\alpha,\beta,\alpha',\beta' \in \C_\text{even}} W_{\alpha,\beta}\left(\Gamma_{\alpha'}^{(1)} \Gamma_{\beta'}^{(2)},\,\left(\frac{i}{2}\gamma^{(1)}_{x,\sigma} \gamma^{(1)}_{x,\tau}\right)\left(-\frac{i}{2}\gamma^{(2)}_{x,\sigma} \gamma^{(2)}_{x,\tau}\right)\Gamma_\alpha^{(1)} \Gamma_\beta^{(2)}\right) \Gamma_{\alpha'}^{(1)} \Gamma_{\beta'}^{(2)} \\
    &= \sum_{\alpha,\beta,\alpha',\beta' \in \C_\text{even}} W_{\alpha,\beta}\left(\Gamma_{\alpha'}^{(1)} ,\,\frac{i}{2}\gamma^{(1)}_{x,\sigma} \gamma^{(1)}_{x,\tau}\Gamma_\alpha^{(1)} \right)\left( \Gamma_{\beta'}^{(2)},\,-\frac{i}{2}\gamma^{(2)}_{x,\sigma} \gamma^{(2)}_{x,\tau} \Gamma_\beta^{(2)}\right) \Gamma_{\alpha'}^{(1)} \Gamma_{\beta'}^{(2)} \\
    &= \sum_{\alpha,\beta,\alpha',\beta' \in \C_\text{even}} W_{\alpha,\beta} (L_{xx,\sigma\tau})_{\alpha',\alpha} (L_{xx,\sigma\tau})_{\beta,\beta'} \Gamma_{\alpha'}^{(1)} \Gamma_{\beta'}^{(2)} \\
    &= \sum_{\alpha,\beta,\alpha',\beta' \in \C_\text{even}}  (L_{xx,\sigma\tau})_{\alpha,\alpha'} W_{\alpha',\beta'}(L_{xx,\sigma\tau})_{\beta',\beta} \Gamma_{\alpha}^{(1)} \Gamma_{\beta}^{(2)}.
  \end{split}
  \label{eq:rep3}
\end{equation}
To Summarize \eqref{eq:rep1}, \eqref{eq:rep2} and \eqref{eq:rep3}, the equation $HO(W)=EO$ is rewritten as
\begin{equation}
  \begin{split}
    &\sum_{x\in A, y \in B}t_{x,y}\sum_{\alpha,\beta,\alpha' \in \C_\text{even}}  \left[(L_{xy,\sigma\sigma})_{\alpha,\alpha'} W_{\alpha',\beta}+W_{\alpha,\alpha'}(L_{xy,\sigma\sigma})_{\alpha',\beta} \right] \Gamma_{\alpha}^{(1)} \Gamma_{\beta}^{(2)} \\
    &+\sum_{x\in \Lambda} U_x\sum_{\alpha,\beta,\alpha',\beta' \in \C_\text{even}}  (L_{xx,\sigma\tau})_{\alpha,\alpha'} W_{\alpha',\beta'}(L_{xx,\sigma\tau})_{\beta',\beta} \Gamma_{\alpha}^{(1)} \Gamma_{\beta}^{(2)}= E \sum_{\alpha,\beta \in \C_\text{even}}   W_{\alpha,\beta} \Gamma_{\alpha}^{(1)} \Gamma_{\beta}^{(2)}.
  \end{split}
\end{equation}
Since each $\Gamma_{\alpha}^{(1)} \Gamma_{\beta}^{(2)}$ is orthogonal, we find
\begin{equation}
  KW+WK+\sum_{x\in \Lambda}\sum_{\sigma,\tau=1}^{N} U_x L_{xx,\sigma \tau} W L_{xx,\sigma \tau} = EW,
\end{equation}
where
\begin{equation}
  (K)_{\alpha,\beta}= \sum_{x\in A, y \in B} \sum_{\sigma=1}^{N}  t_{x,y} (L_{xy,\sigma\sigma})_{\alpha,\beta}.
\end{equation}
The matrix $K$ is Hermitian because $L_{xy,\sigma\sigma}$ is Hermitian and $t_{x,y}$ is real. Similarly, the equation $O(W)H=EO$ is rewritten as
\begin{equation}
  K^{\top}W+WK^{\top}+\sum_{x\in \Lambda}\sum_{\sigma,\tau=1}^{N} U_x L_{xx,\sigma \tau}^{\top} W L_{xx,\sigma \tau}^{\top} = EW.
\end{equation}

\section{Proof of Lemma 6}
Here, we provide a proof of Lemma 6.
Assume that $W$ is a positive semidefinite matrix which satisfies (20) and (21). Suppose that $W$ is not positive definite. Then there exists a vector $\bm{v} \in \Ker W\backslash\{0\}$. Then, by sandwiching (20) by $\bm{v}^\dagger$ and $\bm{v}$, we find
\begin{equation}
  \sum_{x \in \Lambda}\sum_{\sigma,\tau=1}^{N} U_x \bm{v}^\dagger L_{xx,\sigma \tau} W L_{xx,\sigma \tau} \bm{v}=0.
\end{equation}
Since each term in the sum can be rewritten as $\bm{v}^\dagger L_{xx,\sigma \tau} W L_{xx,\sigma \tau} \bm{v}=\left|\sqrt{W} L_{xx,\sigma \tau} \bm{v}\right|^2$,
\begin{equation}
  L_{xx,\sigma \tau} \bm{v} \in \Ker W,\label{Ker1}
\end{equation}
for all $L_{xx,\sigma \tau}$. By operating $\bm{v}$ on (20) from right hand side and using \eqref{Ker1}, we have
\begin{equation}
  K\bm{v} \in \Ker W \label{Ker2}.
\end{equation}
Similarly, from (21), we obtain
\begin{gather}
  L^\top_{xx,\sigma \tau} \bm{v} \in \Ker W,\quad K^\top\bm{v} \in \Ker W \label{Ker3}.
\end{gather}
\par
Next, we identify a vector $\bm{v}\in \mathbb{C}^{|\Ce|}$ with an operator $\Xi=\sum_{\alpha\in \C_\text{even}} v_\alpha \Gamma_\alpha^{(j)}\  (j=1,2)$ such that $(\bm{v})_\alpha=v_\alpha$. Let us choose $j=1$ and define an isomorphism $\varphi$ from a vector to an operator as
\begin{equation}
  \varphi(\bm{v})=\sum_{\alpha\in \C_\text{even}} v_\alpha \Gamma_\alpha^{(1)},\quad \varphi^{-1} \left(\sum_{\alpha\in \C_\text{even}} v_\alpha \Gamma_\alpha^{(1)}\right)=\bm{v}.
\end{equation}
For notational simplicity, we abbreviate $\Gamma_\alpha^{(1)}$ as $\Gamma_\alpha$ and $\gamma^{(1)} _{x,\sigma}$ as  $\gamma _{x,\sigma}$. Let us rewrite \eqref{Ker1}, \eqref{Ker2} and \eqref{Ker3} in terms of operators. Since $(\Gamma_{\alpha},\Gamma_{\beta})=\delta_{\alpha,\beta}$ and $\Gamma_{\alpha}$ is Hermitian, we see that
\begin{equation}
  \begin{split}
    \varphi (L_{xx,\sigma \tau} \bm{v})
    = \sum_{\alpha,\beta\in \C_\text{even}} (L_{xx,\sigma \tau})_{\alpha,\beta}v_\beta \Gamma_\alpha
    = \sum_{\alpha,\beta\in \C_\text{even}} \left(\Gamma_\alpha,\, \frac{i}{2}\gamma_{x,\sigma}\gamma_{x,\tau}\Gamma_\beta\right) v_\beta \Gamma_\alpha
    = \frac{i}{2}\gamma_{x,\sigma}\gamma_{x,\tau}\sum_{\beta\in \C_\text{even}}v_\beta\Gamma_\beta,
  \end{split}
\end{equation}
and
\begin{equation}
  \begin{split}
    \varphi (L^\top_{xx,\sigma \tau} \bm{v})
    = \sum_{\alpha,\beta\in \C_\text{even}} (L_{xx,\sigma \tau})_{\beta,\alpha}v_\beta \Gamma_\alpha
    = \sum_{\alpha,\beta\in \C_\text{even}} \left(\Gamma_\beta,\, \frac{i}{2}\gamma_{x,\sigma}\gamma_{x,\tau}\Gamma_\alpha\right) v_\beta \Gamma_\alpha
    = \left(\sum_{\beta\in \C_\text{even}}v_\beta\Gamma_\beta\right) \frac{i}{2}\gamma_{x,\sigma}\gamma_{x,\tau}.
  \end{split}
\end{equation}
Similarly,
\begin{gather}
  \varphi (K\bm{v})= \left[\sum_{x\in A, y \in B} \sum_{\sigma=1}^{N}\frac{i}{2}t_{x,y}\gamma_{x,\sigma}\gamma_{y,\sigma}\right]\sum_{\beta\in \C_\text{even}}v_\beta\Gamma_\beta,\\
  \varphi (K^\top \bm{v})=\sum_{\beta\in \C_\text{even}}v_\beta\Gamma_\beta \left[\sum_{x\in A, y \in B} \sum_{\sigma=1}^{N}\frac{i}{2}t_{x,y}\gamma_{x,\sigma}\gamma_{y,\sigma}\right].
\end{gather}
From \eqref{Ker1}, \eqref{Ker2} and \eqref{Ker3}, we see that, if $\varphi^{-1} (\Xi) \in \Ker W$,
\begin{align}
  \varphi^{-1} (\gamma_{x,\sigma}\gamma_{x,\tau}\Xi)&,\  \varphi^{-1} (\Xi\gamma_{x,\sigma}\gamma_{x,\tau})\in \Ker W \label{opKer1}, \\
  \varphi^{-1}\left(\tilde{K}\Xi\right)&,\  \varphi^{-1}\left(\Xi\tilde{K} \right)\in \Ker W, \label{opKer2}
\end{align}
where $\tilde{K}=\left[\sum_{x\in A, y \in B} \sum_{\sigma=1}^{N}t_{x,y}\gamma_{x,\sigma}\gamma_{y,\sigma}\right]$. The following lemma follows from \eqref{opKer1} and \eqref{opKer2}.
\par
{\it Lemma 9.}---
Let $W$ be a positive semidefinite matrix which satisfies (20) and (21). Assume that $N\geq 3$. If $\varphi^{-1} (\Xi) \in \Ker W$, then
\begin{equation}
  \varphi^{-1}(\gamma_{x,\sigma}\gamma_{y,\tau}\Xi),\quad \varphi^{-1}(\Xi\gamma_{x,\sigma}\gamma_{y,\tau}) \in \Ker W,
  \label{eq:anyhopKer}
\end{equation}
for all $(x,\sigma),\,(y,\tau) \in \Lambda\times \{1,\cdots N\}$.
\par
{\it Proof of Lemma 9.}---
Let us write as $\mathscr{N}(x)$ the subset of $\Lambda$ which is directly connected with a site $x\in\Lambda$ through a nonvanishing hopping matrix element. If $x$ belongs to the sublattice $A$,
\begin{equation}
  \begin{split}
    [\gamma_{x,\sigma}\gamma_{x,\tau},\tilde{K}]
    =\sum_{y\in\mathscr{N}(x)}t_{x,y}[\gamma_{x,\sigma}\gamma_{x,\tau},(\gamma_{x,\sigma}\gamma_{y,\sigma}+\gamma_{x,\tau}\gamma_{y,\tau})]
    =2\sum_{y\in\mathscr{N}(x)}t_{x,y}(\gamma_{x,\sigma}\gamma_{y,\tau}-\gamma_{x,\tau}\gamma_{y,\sigma} ).
  \end{split}
\end{equation}
Since $N\geq3$, we can take a flavor $\upsilon$ which is different from $\sigma$ and $\tau$. Then, for a site $z\in \mathscr{N}(x)$,
\begin{equation}
  \begin{split}
    \bigl[\gamma_{z,\upsilon}\gamma_{z,\sigma},[\gamma_{x,\sigma}\gamma_{x,\tau},\tilde{K}]\bigr]
    =2\sum_{y\in\mathscr{N}(x)}t_{x,y}\bigl[\gamma_{z,\upsilon}\gamma_{z,\sigma},(\gamma_{x,\sigma}\gamma_{y,\tau}-\gamma_{x,\tau}\gamma_{y,\sigma})\bigr]
    =4t_{x,z}\gamma_{z,\upsilon}\gamma_{x,\tau}.
  \end{split}
\end{equation}
Similarly, if $x$ belongs to the sublattice $B$,
\begin{equation}
  \bigl[\gamma_{z,\upsilon}\gamma_{z,\sigma},[\gamma_{x,\sigma}\gamma_{x,\tau},\tilde{K}]\bigr]=-4t_{z,x}\gamma_{z,\upsilon}\gamma_{x,\tau}.
\end{equation}
From \eqref{opKer1} and \eqref{opKer2},
\begin{equation}
  \varphi^{-1}\left(  \bigl[\gamma_{z,\upsilon}\gamma_{z,\sigma},[\gamma_{x,\sigma}\gamma_{x,\tau},\tilde{K}]\bigr]\Xi\right)
  ,\quad\varphi^{-1}\left( \Xi \bigl[\gamma_{z,\upsilon}\gamma_{z,\sigma},[\gamma_{x,\sigma}\gamma_{x,\tau},\tilde{K}]\bigr]\right)
  \in \Ker W.
\end{equation}
Since $t_{x,z}$ and  $t_{z,x}$ are nonvanishing,
\begin{equation}
  \varphi^{-1}(\gamma_{x,\tau}\gamma_{z,\upsilon}\Xi),\quad \varphi^{-1}(\Xi\gamma_{x,\tau}\gamma_{z,\upsilon}) \in \Ker W.
  \label{eq:newhopKer}
\end{equation}
To summarize \eqref{opKer1} and \eqref{eq:newhopKer}, we have shown that
$\varphi^{-1}(\gamma_{x,\sigma}\gamma_{y,\tau}\Xi),\  \varphi^{-1}(\Xi\gamma_{x,\sigma}\gamma_{y,\tau}) \in \Ker W$ for all \newline${(x,\sigma),\,(y,\tau) \in \Lambda\times \{1,\cdots N\}}$ which satisfies
\begin{equation}
  x=y,\ \text{or}\  (t_{x,y}\neq0\  \text{and}\  \sigma\neq \tau).
  \label{eq:condition}
\end{equation}
\par
Consider arbitrary two pairs of a site and a flavor ${(x,\sigma),\, (y,\tau) \in \Lambda\times \{1,\cdots N\}}$. Since we have assumed that the lattice $\Lambda$ is connected via nonvanishing hopping matrix elements, there exists a finite sequence of sites and flavors ${(z_1,\upsilon_1),\cdots,(z_n,\upsilon_m) \in \Lambda\times \{1,\cdots N\}}$, such that $(z_1,\upsilon_1)=(x,\sigma)$, $(z_m,\upsilon_n)=(y,\tau)$ and all neighboring pairs $\{(z_j,\upsilon_j), (z_{j+1},\upsilon_{j+1})\}\,(j=1,\cdots m-1)$ satisfy the condition \eqref{eq:condition}. Noting that
\begin{equation}
  \prod_{j=1}^{m-1}\gamma_{z_j,\upsilon_j}\gamma_{z_{j+1},\upsilon_{j+1}}= \gamma_{x,\sigma}\gamma_{y,\tau},
\end{equation}
where the product on left hand side is ordered in ascending order of $j$, we find that $\varphi^{-1}(\gamma_{x,\sigma}\gamma_{y,\tau}\Xi),\ \varphi^{-1}(\Xi\gamma_{x,\sigma}\gamma_{y,\tau}) \in \Ker W$ for an arbitrary pair $(x,\sigma),\, (y,\tau) \in \Lambda\times \{1,\cdots N\}$.
\hspace{\fill}$\blacksquare$
\par
Consider a monomial $\Gamma_\alpha$ which contains even number of Majorana operator. Then,
\begin{equation}
  \gamma_{x,\sigma} \gamma_{y,\tau} \Gamma_\alpha = \left\{
  \begin{array}{cl}
    \Gamma_\alpha \gamma_{x,\sigma} \gamma_{y,\tau} &
    \mathrm{if}\ (x,\sigma),(y,\tau)\in \alpha \ \mathrm{or}\ (x,\sigma),(y,\tau)\notin \alpha, \\
    -\Gamma_\alpha \gamma_{x,\sigma} \gamma_{y,\tau} & \mathrm{otherwise}.
  \end{array}
  \right.
\end{equation}
From the above relations, we can define a projection acting on $\Xi=\sum_{\alpha\in \C_{\text{even}}}$ $v_\alpha \Gamma_\alpha$,
\begin{equation}
  P_{(x,\sigma),(y,\tau)} \Xi= \frac{1}{2} \left[\gamma_{y,\tau}\gamma_{x,\sigma}\Xi\gamma_{x,\sigma}\gamma_{y,\tau}+\Xi\right].
\end{equation}
This is a projection to a space spanned by $\{\Gamma_\alpha\,|\alpha\in \C_\text{even},\,(x,\sigma),(y,\tau)\in \alpha\ \text{or}\ (x,\sigma),(y,\tau)\notin \alpha\}$. From Lemma 9, if $\varphi^{-1}(\Xi)\in \Ker W$,
\begin{gather}
  \varphi^{-1}(P_{(x,\sigma),(y,\tau)}\Xi)\in \Ker W,
  \label{eq:projection}
\end{gather}
for all $(x,\sigma),\,(y,\tau) \in \Lambda\times \{1,\cdots N\}$. Now we are ready to prove Lemma 6.
\par
{\it Proof of Lemma 6.}---
Let $W$ be a positive semidefinite matrix which satisfies (20) and (21). Suppose that $W$ is not positive definite. Then, there exists a vector $\bm{v} \in \Ker W\backslash \{0\}$. Define a corresponding operator as  $\Xi=\varphi(\bm{v})=\sum_{\alpha\in \C_{\text{even}}}$ $v_\alpha \Gamma_\alpha$. Let $v_\beta$ be a nonzero component of $\bm{v}$. By successively operating $\gamma_{x,\sigma}\gamma_{y,\tau}$, where $(x,\sigma),\,(y,\tau) \in \Lambda\times \{1,\cdots N\}$, we can bring $\Gamma_\beta$ to $\Gamma_0=1$, since the length of $\beta$ is even. Then we obtain a new vector $\bm{v}'\in \Ker W\backslash \{0\}$ such that $v_0'\neq 0$. Here, $v_0'$ is the component of $\bm{v}'$ corresponding to $\Gamma_0$.
\par
Next, we use \eqref{eq:projection} to $\Xi'=\varphi(\bm{v}')$.
If $NN_{\mathrm{s}}$ is odd, by successively applying projections to $\Xi'$, we can drop all the terms other than $v'_0\Gamma_0$. Then, we find that $\varphi^{-1}(\Gamma_0)\in \Ker W$.
\par
Finally, by successively operating $\gamma_{x,\sigma}\gamma_{y,\tau}$, we can bring $\Gamma_0$ to $\Gamma_\alpha$ for any $\alpha \in \Ce$. Thus $\varphi^{-1}\left(  \Gamma_\alpha\right)\in \Ker W$. Therefore, for any vector $\bm{w} \in \mathbb{C}^{|\Ce|}$,
\begin{equation}
  \bm{w} = \varphi^{-1}\left(\sum_{\alpha\in \C_{\text{even}}} w_\alpha \Gamma_\alpha\right)
  = \sum_{\alpha\in \C_{\text{even}}} w_\alpha \varphi^{-1}\left(  \Gamma_\alpha\right)\in \Ker W,
\end{equation}
which means that $W$ is zero.
\hspace{\fill}$\blacksquare$

\section{The ground states in the large-$U_x$ limit}

In the the large-$U_x$ limit, the effective Hamiltonian of the one-dimensional attractive SU($N$) model with $N\geq3$  is derived in Ref. \cite{Zhao2007}. Here we generalize it to general bipartite lattices and show that the ground states correspond to (26). In the large-$U_x$ limit, we treat $H_\mathrm{int}$ as the unperturbed Hamiltonian and the hopping Hamiltonian $H_\mathrm{hop}$ as a perturbation.
\begin{align}
  H_{\mathrm{int}}&=\sum_{x\in \Lambda} U_x n_x\left(n_x-N\right), \\
  H_{\mathrm{hop}}&=\sum_{x\in A, y \in B} \sum_{\sigma=1}^{N} t_{x,y} (c^\dagger_{x,\sigma} c_{y,\sigma}+c^\dagger_{y,\sigma} c_{x,\sigma}).
\end{align}
where we shifted $H_{\mathrm{int}}$ by a constant to make the ground state energy zero. Note that $U_x<0$.
\par
We first consider the unperturbed Hamiltonian $H_\mathrm{int}$. Then the number operator $n_{x}$ at each site $x\in\Lambda$ commutes with the Hamiltonian. The energy is minimized when $n_x = 0 \ \mathrm{or}\  N$ for all $x\in \Lambda$. Thus the ground states are $2^{N_{\mathrm{s}}}$-fold degenerate. We define a new classical variable $s_x$ on each site to write down the effective Hamiltonian.
\begin{equation}
  s_x = \left\{
  \begin{array}{cc}
    1 &
    \mathrm{if}\ n_x = N, \\
    -1 & \mathrm{if}\ n_x = 0.
  \end{array}
  \right.
\end{equation}
\par
Let us examine the effect of the perturbation $H_\mathrm{hop}$. Since the first order vanishes, we consider the second order perturbation. Let $P_0$ be the orthogonal projection to the ground states of $H_{\mathrm{int}}$. Then, the low energy effective Hamiltonian is
\begin{equation}
  H_\mathrm{eff}=-P_0 H_{\mathrm{hop}} H_{\mathrm{int}}^{-1} H_{\mathrm{hop}} P_0.
\end{equation}
Let $\ket{\psi}$ be a ground state of $H_{\mathrm{int}}$. Then,
\begin{equation}
  H_{\mathrm{int}}^{-1} H_{\mathrm{hop}} \ket{\psi}= \sum_{x\in A, y \in B} \sum_{\sigma=1}^{N} \frac{t_{x,y}}{(N-1)\left(|U_x|+|U_y|\right)} (c^\dagger_{x,\sigma} c_{y,\sigma}+c^\dagger_{y,\sigma} c_{x,\sigma})\ket{\psi},
\end{equation}
because $c_{x(y),\sigma}^\dagger c_{y(x),\sigma}\ket{\psi}$ is a state with one fermion at site $x\,(y)$ and $N-1$ fermions at site $y\,(x)$ if it is nonzero, and
\begin{equation}
  H_{\mathrm{int}}c_{x(y),\sigma}^\dagger c_{y(x),\sigma}\ket{\psi}=(N-1)(|U_x|+|U_y|)\ket{\psi}.
\end{equation}
One thus finds
\begin{equation}
  \begin{split}
    P_0 H_{\mathrm{hop}} H_{\mathrm{int}}^{-1} H_{\mathrm{hop}} P_0
    &= P_0\sum_{x\in A, y \in B} \sum_{\sigma=1}^{N} \frac{t_{x,y}^2}{(N-1)\left(|U_x|+|U_y|\right)} (c^\dagger_{x,\sigma} c_{y,\sigma}+c^\dagger_{y,\sigma} c_{x,\sigma})(c^\dagger_{x,\sigma} c_{y,\sigma}+c^\dagger_{y,\sigma} c_{x,\sigma}) P_0\\
    &=P_0\sum_{x\in A, y \in B} \sum_{\sigma=1}^{N} \frac{t_{x,y}^2}{(N-1)\left(|U_x|+|U_y|\right)} [n_{x,\sigma}(1-n_{y,\sigma})+n_{y,\sigma}(1-n_{x,\sigma})]P_0.
  \end{split}
\end{equation}
If $n_x=n_y=0\,(N)$, then $n_{x,\sigma}=n_{y,\sigma}=0\,(1)$. Thus
\begin{equation}
  n_{x,\sigma}(1-n_{y,\sigma})+n_{y,\sigma}(1-n_{x,\sigma})= 0.
\end{equation}
If $n_x= 0\,(N)$ and $n_y=N\,(0)$, then $n_{x,\sigma}= 0\,(1)$ and $n_{y,\sigma}= 1\,(0)$. Thus
\begin{equation}
  n_{x,\sigma}(1-n_{y,\sigma})+n_{y,\sigma}(1-n_{x,\sigma})= 1.
\end{equation}
Therefore, the effective Hamiltonian reads
\begin{equation}
  H_\mathrm{eff}=\sum_{x\in A, y \in B}\frac{Nt_{x,y}^2}{(N-1)\left(|U_x|+|U_y|\right)}\left(\frac{s_x s_y-1}{2}\right),
\end{equation}
if restricted to the ground states of the unperturbed Hamiltonian $H_{\mathrm{int}}$. The effective model turns out to be the antiferromagnetic Ising model on $\Lambda$. This is qualitatively understood as follows: the energy of each $N$-fermion bound state is decreased by quantum fluctuations where one of the fermions virtually hops to one of the neighboring sites. These quantum fluctuations are reduced if $N$-fermion bound states sit next to each other. In the ground states of the effective Hamiltonian,
\begin{equation}
  s_x=\left\{
  \begin{array}{cc}
    1 &
    \mathrm{if}\ x \in A, \\
    -1 &
    \mathrm{if}\ x \in B,
  \end{array}
  \right.
  \quad \mathrm{or} \quad
  s_x=\left\{
  \begin{array}{cc}
    -1 &
    \mathrm{if}\ x \in A, \\
    1 &
    \mathrm{if}\ x \in B.
  \end{array}
  \right.
  \label{eq:gs_of_effective_model}
\end{equation}
These states correspond to (26).
\section{Proof of Theorems 1 and 2 for even $NN_{\mathrm{s}}$}
Here we provide proofs of Theorems 1 and 2 for even $NN_{\mathrm{s}}$. When $NN_{\mathrm{s}}$ is even, two parity operators $\Delta^{(1)}$ and $\Delta^{(2)}$ commute. Since $H$ also commutes with $\Delta^{(1)}$ and $\Delta^{(2)}$, $H$ is block-diagonal in the basis where $\Delta^{(1)}$ and $\Delta^{(2)}$ are diagonal. Note that the eigenvalues of $\Delta^{(1)}$ and $\Delta^{(2)}$ are $\pm1$. Then the eigenoperators can be classified into four sectors as
\begin{equation}
  O \in \mathscr{O}^{p_1 p_2}\quad \Leftrightarrow\quad \Delta_1 O=O\Delta_1=p_1O,\quad \Delta_2 O=O\Delta_2=p_2O\quad (p_1,p_2=\pm).
  \label{eq:pmsectors2}
\end{equation}
The following lemma shows that all the eigenoperators are in $\mathscr{O}_{\mathrm{even,even}}$ sector.
\par
{\it Lemma 10.}---
Let $\mathscr{O}^{p_1 p_2}$ $(p_1,p_2=\pm)$ be subsets of $\mathscr{O}$ defined by \eqref{eq:pmsectors2}. Then,
\begin{equation}
  \mathscr{O}^{++}
  \oplus \mathscr{O}^{+-}
  \oplus \mathscr{O}^{-+}
  \oplus \mathscr{O}^{--}  =\mathscr{O}_{\mathrm{even,even}}.
  \label{eq:pmsectors1}
\end{equation}
{\it Proof of Lemma 10.}---
By definition, all the elements of $\mathscr{O}^{p_1 p_2}$ commute with $\Delta^{(1)}$ and $\Delta^{(2)}$. On the other hand, a basis operator $\Gamma^{(j)}_\alpha$ $(j=1,2)$ commutes (anticommutes) with $\Delta^{(j)}$ when the length of $\alpha$ is even (odd). Thus ${\mathscr{O}^{++}
\oplus \mathscr{O}^{+-}
\oplus \mathscr{O}^{-+}
\oplus \mathscr{O}^{--} \subset \mathscr{O}_\text{even,even}}$.
\par
For any even-length configuration $\alpha\in \Ce$, we can symmetrize or antisymmetrize $\Gamma^{(j)}_{\alpha}$ and $\Delta^{(j)}\Gamma^{(j)}_{\alpha}$ as
\begin{equation}
  \Gamma^{(j,\pm)}_{\alpha}= \frac{1}{\sqrt{2}}[1\pm \Delta^{(j)}]\Gamma_\alpha^{(j)},
  \label{eq:defgammapm}
\end{equation}
Note that $\left(\Gamma^{(j,p_1)}_{\alpha},\Gamma^{(k,p_2)}_{\alpha}\right)=\delta_{j,k}\delta_{p_1,p_2}$. Since $\Gamma^{(1,p_1)}_{\alpha}\Gamma^{(2,p_2)}_{\beta}$ is in $\mathscr{O}^{p_1 p_2}$ sector, ${\mathscr{O}^{++} \oplus \mathscr{O}^{+-} \oplus \mathscr{O}^{-+} \oplus \mathscr{O}^{--}} \supset \mathscr{O}_\text{even,even}$. Therefore, we obtain ${\mathscr{O}^{++} \oplus \mathscr{O}^{+-} \oplus \mathscr{O}^{-+} \oplus \mathscr{O}^{--}}  =\mathscr{O}_{\mathrm{even,even}}$.
\hspace{\fill}$\blacksquare$
\par
For an even-length configuration $\alpha \in \Ce$, we define a configuration $\overline{\alpha}\in \C_\text{even}$ which satisfy $\Gamma^{(j)}_{\overline{\alpha}}=\pm
\Delta^{(j)}\Gamma_\alpha^{(j)}$, where $\pm$ comes from the anticommutativity of Majorana operators. We identify $\overline{\alpha}$ with $\alpha$ and write the quotient set of $\C_\text{even}$ by this identification as $\C'_\text{even}$.
\par
For $p_1,p_2=\pm$, an operator in each sector $\mathscr{O}^{p_1 p_2}$ is expressed as
\begin{equation}
  O^{p_1 p_2}(W)=\sum_{\alpha,\beta \in \C'_\text{even}} W_{\alpha,\beta} \Gamma^{(1,p_1)}_\alpha\Gamma^{(2,p_2)}_\beta.
  \label{eq:pm_expansion}
\end{equation}
Like in the odd $NN_{\mathrm{s}}$ case, the eigenequations read as follows.
\begin{equation}
  K^{p_1}W+WK^{p_2}+\sum_{x\in \Lambda}\sum_{\sigma,\tau=1}^{N} U_x L^{p_1}_{xx,\sigma \tau} W L^{p_2}_{xx,\sigma \tau} = EW,
  \label{eq:eveneigen1}
\end{equation}
\begin{equation}
  K^{p_1\top}W+WK^{p_2\top}+\sum_{x\in \Lambda}\sum_{\sigma,\tau=1}^{N} U_x L_{xx,\sigma \tau}^{p_1\top} W L_{xx,\sigma \tau}^{p_2\top} = EW,
  \label{eq:eveneigen2}
\end{equation}
where $L^{p_1}_{xy,\sigma\tau}$ and $K^{p_1}$ are $|\C'_\text{even}|\times |\C'_\text{even}|$ Hermitian matrices defined by
\begin{gather}
  (L^{p_1}_{xy,\sigma\tau})_{\alpha,\beta}=\left(\Gamma^{(1,p_1)}_\alpha,\, \frac{i}{2}\gamma_{x,\sigma}^{(1)}\gamma_{y,\tau}^{(1)}\Gamma^{(1,p_1)}_\beta\right), \label{eq:def_of_L_pm}\\
  (K^{p_1})_{\alpha,\beta}= \sum_{x\in A, y \in B} \sum_{\sigma=1}^{N}  t_{x,y} (L^{p_1}_{xy,\sigma\sigma})_{\alpha,\beta}\label{eq:def_of_K_pm}.
\end{gather}
Next, we consider the expectation value of the Hamiltonian $H$. Since $(O^{p_1 p_2}(W),O^{p_1 p_2}(W))=\Tr[W^\dagger W]$, the normalization condition for operators $(O,O)=1$ can be rewritten as $\Tr[W^2]=1$ with a $|\C'_\text{even}|\times |\C'_\text{even}|$ Hermitian matrix $W$. We define $E^{p_1, p_2}(W)=\left(O^{p_1 p_2}(W),H O^{p_1 p_2}(W)\right)$ for a normalized Hermitian matrix $W$. Then, $E^{p_1, p_2}(W)$ is calculated as
\begin{equation}
  E^{p_1, p_2}(W)= \Tr \left[\left(K^{p_1}+K^{p_2}\right) W^2\right]+ \sum_{x\in \Lambda}\sum_{\sigma,\tau=1}^{N} U_x \Tr\left[ WL^{p_1}_{xx,\sigma \tau} W L^{p_2}_{xx,\sigma \tau}\right].
  \label{eq:evenexpectation}
\end{equation}
\par
Within each $\mathscr{O}^{p_1 p_2}$ sector, we can show that the lowest energy eigenoperator is unique in much the same way as in the case of $N N_{\mathrm s}$ odd (The small difference is that we have identified $\alpha \in \C_\text{even}$ and $\overline{\alpha}\in \C_\text{even}$).
\par
{\it Lemma 11.}---
Within each $\mathscr{O}^{p_1 p_2}$ $(p_1,p_2=\pm)$ sector, the lowest energy eigenoperator is unique. If the lowest energy eigenoperator in each sector is expressed as $O^{p_1 p_2}(W)$, then $W$ is either positive or negative definite.
\par
We define the lowest energy in $\mathscr{O}^{p_1 p_2}$ sector as $E^{p_1 p_2}_{\mathrm{GS}}$.
Then the following lemma shows that the degeneracy of the ground states is at most two in total. Thus Lemma 4 was proved.
\par
{\it Lemma 12.}---
Consider the lowest energy in each sector, $E^{p_1 p_2}_{\mathrm{GS}}$ $(p_1,p_2=\pm)$. Then,
\begin{equation}
  E^{+ -}_{\mathrm{GS}}=E^{- +}_{\mathrm{GS}}> \frac{1}{2}\left(E^{+ +}_{\mathrm{GS}}+E^{- -}_{\mathrm{GS}}\right).
  \label{eq:pmgs}
\end{equation}
\par
{\it Proof of Lemma 12.}---
Let the lowest energy eigenoperator in $\mathscr{O}^{+-}$ sector be $O^{+-}(W)$, where $W$ is a $|\C'_\text{even}| \times |\C'_\text{even}|$ positive or negative definite matrix. See \eqref{eq:pm_expansion} for the matrix representation of operators in each sector. By \eqref{eq:evenexpectation}, $E^{- +}(W)=E^{+ -}(W)=E^{+ -}_{\mathrm{GS}}$, and hence $E^{- +}_{\mathrm{GS}} \leq E^{+ -}_{\mathrm{GS}}$. Since we can similarly show that $E^{+ -}_{\mathrm{GS}} \leq E^{- +}_{\mathrm{GS}}$, we obtain $E^{+-}_{\mathrm{GS}} = E^{-+}_{\mathrm{GS}}$ and $O^{-+}(W)$ is the lowest energy eigenoperator in $\mathscr{O}^{-+}$ sector.
\par
For the same matrix $W$, we consider operators $O^{+ +}(W)\in\mathscr{O}^{++}$ and  $O^{- -}(W)\in\mathscr{O}^{--}$. We would like to compare the energy expectation values of $O^{+ +}(W)$, $O^{+ -}(W)$, $O^{- +}(W)$ and $O^{- -}(W)$. Using \eqref{eq:evenexpectation}, one finds
\begin{equation}
  \begin{split}
    &\Bigl[E^{+-}(W)+E^{-+}(W)\Bigr]-\Bigl[E^{++}(W)+E^{--}(W)\Bigr] \\
    &=\Tr \left[\left(K^{+}+K^{-}\right) W^2\right]+ \sum_{x\in \Lambda}\sum_{\sigma,\tau=1}^{N} U_x \Tr\left[ W L^{+}_{xx,\sigma \tau} W L^{-}_{xx,\sigma \tau}\right] \\
    &+ \Tr \left[\left(K^{-}+K^{+}\right) W^2\right]+ \sum_{x\in \Lambda}\sum_{\sigma,\tau=1}^{N} U_x \Tr\left[ W L^{-}_{xx,\sigma \tau} W L^{+}_{xx,\sigma \tau}\right] \\
    &- \Tr \left[\left(K^{+}+K^{+}\right) W^2\right]- \sum_{x\in \Lambda}\sum_{\sigma,\tau=1}^{N} U_x \Tr\left[ W L^{+}_{xx,\sigma \tau} W L^{+}_{xx,\sigma \tau}\right] \\
    &- \Tr \left[\left(K^{-}+K^{-}\right) W^2\right]- \sum_{x\in \Lambda}\sum_{\sigma,\tau=1}^{N} U_x \Tr\left[ W L^{-}_{xx,\sigma \tau} W L^{-}_{xx,\sigma \tau}\right] \\
    &=-\sum_{x\in \Lambda}\sum_{\sigma,\tau=1}^{N}U_x \Tr [W(L^{+}_{xx,\sigma\tau}-L^{-}_{xx,\sigma\tau}) W(L^{+}_{xx,\sigma\tau}-L^{-}_{xx,\sigma\tau})].
    \label{eq:energydiff}
  \end{split}
\end{equation}
The matrix elements of $L^{\pm}_{xx,\sigma\tau}$ are calculated as
\begin{equation}
  \begin{split}
    (L^{+}_{xx,\sigma\tau})_{\alpha,\beta}
    &=\frac{1}{2} \left(\left[1+\Delta^{(1)}\right]\Gamma_\alpha^{(1)},\, \frac{i}{2}\gamma_{x,\sigma}^{(1)}\gamma_{y,\tau}^{(1)}\left[1+\Delta^{(1)}\right]\Gamma_\beta^{(1)}\right)\\
    &=\frac{i}{4} \left(\Gamma_\alpha^{(1)},\, \gamma_{x,\sigma}^{(1)}\gamma_{y,\tau}^{(1)}\Gamma_\beta^{(1)}\right)+\frac{i}{4} \left(\Delta^{(1)}\Gamma_\alpha^{(1)},\, \gamma_{x,\sigma}^{(1)}\gamma_{y,\tau}^{(1)}\Delta^{(1)}\Gamma_\beta^{(1)}\right) \\
    &+\frac{i}{4} \left(\Gamma_\alpha^{(1)},\, \gamma_{x,\sigma}^{(1)}\gamma_{y,\tau}^{(1)}\Delta^{(1)}\Gamma_\beta^{(1)}\right)+\frac{i}{4} \left(\Delta^{(1)}\Gamma_\alpha^{(1)},\, \gamma_{x,\sigma}^{(1)}\gamma_{y,\tau}^{(1)}\Gamma_\beta^{(1)}\right),
  \end{split}
\end{equation}
\begin{equation}
  \begin{split}
    (L^{-}_{xx,\sigma\tau})_{\alpha,\beta}
    &=\frac{1}{2} \left(\left[1-\Delta^{(1)}\right]\Gamma_\alpha^{(1)},\, \frac{i}{2}\gamma_{x,\sigma}^{(1)}\gamma_{y,\tau}^{(1)}\left[1-\Delta^{(1)}\right]\Gamma_\beta^{(1)}\right)\\
    &=\frac{i}{4} \left(\Gamma_\alpha^{(1)},\, \gamma_{x,\sigma}^{(1)}\gamma_{y,\tau}^{(1)}\Gamma_\beta^{(1)}\right)+\frac{i}{4} \left(\Delta^{(1)}\Gamma_\alpha^{(1)},\, \gamma_{x,\sigma}^{(1)}\gamma_{y,\tau}^{(1)}\Delta^{(1)}\Gamma_\beta^{(1)}\right) \\
    &-\frac{i}{4} \left(\Gamma_\alpha^{(1)},\, \gamma_{x,\sigma}^{(1)}\gamma_{y,\tau}^{(1)}\Delta^{(1)}\Gamma_\beta^{(1)}\right)-\frac{i}{4} \left(\Delta^{(1)}\Gamma_\alpha^{(1)},\, \gamma_{x,\sigma}^{(1)}\gamma_{y,\tau}^{(1)}\Gamma_\beta^{(1)}\right).
  \end{split}
\end{equation}
Then we can calculate $L^{+}_{xx,\sigma\tau}-L^{-}_{xx,\sigma\tau}$ as
\begin{equation}
  \begin{split}
    (L^{+}_{xx,\sigma\tau}-L^{-}_{xx,\sigma\tau})_{\alpha,\beta}
    &= \frac{i}{2} \left(\Gamma_\alpha^{(1)},\, \gamma_{x,\sigma}^{(1)}\gamma_{y,\tau}^{(1)}\Delta^{(1)}\Gamma_\beta^{(1)}\right)+\frac{i}{2} \left(\Delta^{(1)}\Gamma_\alpha^{(1)},\, \gamma_{x,\sigma}^{(1)}\gamma_{y,\tau}^{(1)}\Gamma_\beta^{(1)}\right)\\
    &= i \left(\Gamma_\alpha^{(1)},\, \gamma_{x,\sigma}^{(1)}\gamma_{y,\tau}^{(1)}\Delta^{(1)}\Gamma_\beta^{(1)}\right)
    = \pm i \left(\Gamma_\alpha^{(1)},\, \gamma_{x,\sigma}^{(1)}\gamma_{y,\tau}^{(1)}\Gamma_{\overline{\beta}}^{(1)}\right),
  \end{split}
  \label{eq:lpluslminus}
\end{equation}
where we noted that $\Delta^{(1)}$ commutes with $\Gamma_\alpha^{(1)}$ and $\Gamma_\beta^{(1)}$ to get the second line. From \eqref{eq:lpluslminus}, ${L^{+}_{xx,\sigma\tau}-L^{-}_{xx,\sigma\tau}}$ is Hermitian and nonvanishing. This, together with Lemma 8, implies that ${\Tr [W(L^{+}_{xx,\sigma\tau}-L^{-}_{xx,\sigma\tau}) W(L^{+}_{xx,\sigma\tau}-L^{-}_{xx,\sigma\tau})]}>0$. Therefore, from \eqref{eq:energydiff}, ${\Bigl[E^{+-}(W)+E^{-+}(W)\Bigr]-\Bigl[E^{++}(W)+E^{--}(W)\Bigr]}>0$. Since $E^{+-}(W)=E^{+-}_{\mathrm{GS}}$, $E^{-+}(W)=E^{-+}_{\mathrm{GS}}$, $E^{++}(W)\geq E^{++}_{\mathrm{GS}}$, $E^{--}(W)\geq E^{--}_{\mathrm{GS}}$ and $E^{+-}_{\mathrm{GS}}=E^{-+}_{\mathrm{GS}}$, we obtain \eqref{eq:pmgs}.
\hspace{\fill}$\blacksquare$
\par
Here we have completed the proof of Lemma 4. For even $NN_{\mathrm{s}}$, Theorem 1 can be derived from Lemmas 4 and 11.
\par
{\it Proof of Theorem 1 for even $NN_{\mathrm{s}}$.}---
Because the ground state degeneracy is at most two, any ground state must be an SU($N$) singlet. To determine the fermion number, we consider the states $\ket{\Phi_{+}}$ and $\ket{\Phi_{-}}$ characterized by (26), whose fermion numbers are $NN_A$ and $NN_B$, respectively. The eigenoperors are written as
$O_\pm = \ket{\Phi_{\pm}}\bra{\Phi_{\pm}} = 2^{-NN_\mathrm{s}}\prod_{x\in \Lambda}\prod_{\sigma=1}^N \left(1\pm i\gamma^{(1)}_{x,\sigma}\gamma^{(2)}_{x,\sigma}\right)$.
Then, $2^{NN_\mathrm{s}-1} (O_{+}+O_{-})$ is in $\mathscr{O}_\mathrm{even,even}$ sector and written as $O(I)$, where $I$ is the identity matrix of size $|\Ce|$. $O(I)$ can be expanded as
\begin{equation}
  \begin{split}
      O(I)&=\frac{(1+\Delta^{(1)})}{2}\frac{(1+\Delta^{(2)})}{2}O(I)+\frac{(1+\Delta^{(1)})}{2}\frac{(1-\Delta^{(2)})}{2}O(I) \\
      &+\frac{(1-\Delta^{(1)})}{2}\frac{(1+\Delta^{(2)})}{2}O(I)+\frac{(1-\Delta^{(1)})}{2}\frac{(1-\Delta^{(2)})}{2}O(I).
  \end{split}
  \label{eq:identity_pm}
\end{equation}
Note that $(1 \pm \Delta^{(1)})/2 \times(1 \pm \Delta^{(2)})/2$ is a projection operator to $\mathcal{O}^{\pm \pm}$ sector. With the matrix expression \eqref{eq:pm_expansion}, \eqref{eq:identity_pm} is rewritten as
\begin{equation}
  O(I)=O^{++}\left(I'\right)+O^{+-}\left(I'\right)+O^{-+}\left(I'\right)+O^{--}\left(I'\right),
\end{equation}
where $I'$ is the identity matrix of size $|\Ce '|$. Let $O^{++}(W_{\mathrm{GS}})$ be the lowest energy eigenoperator in $\mathscr{O}^{+ +}$ sector. By Lemma 11,
$\left(O(I),O^{++}\left(W_\mathrm{GS}\right)\right)=\Tr\left[{W_\mathrm{GS}}\right]\neq0$, because $W_\mathrm{GS}$ is positive or negative definite. As in the case of $NN_{\mathrm{s}}$ odd, the fermion number of the lowest energy state in $\mathscr{O}^{+ +}$ sector is $NN_A$ or $NN_B$. Similarly, the fermion number of the lowest energy state in $\mathscr{O}^{--}$ sector is $NN_A$ or $NN_B$. If $N_A\neq N_B$, by Lemma 4 and the particle-hole symmetry of the model, there are exactly two ground states and the fermion numbers are $NN_A$ and $NN_B$, respectively. If $N_A$=$N_B$, by Lemma 4, there are at most two ground states and whose total fermion number is $NN_A$ $(=NN_B)$.
\hspace{\fill}$\blacksquare$
\par
Finally, we prove Theorem 2 for even $NN_{\mathrm{s}}$.
\par
{\it Proof of Theorem 2 for even $NN_{\mathrm{s}}$.}---
When $NN_{\mathrm{s}}$ is even, the lowest state eigenoperators in each $\mathscr{O}^{p_1 p_2}$ sector (See \eqref{eq:pmsectors2} for the definition) is unique.  We already know that the ground state eigenoperator is in $\mathscr{O}^{+ +} \oplus \mathscr{O}^{- -}$ sector. Assume that the ground state eigenoperator is in $\mathscr{O}^{+ +}$ sector and expressed as $O^{+ +}(W_\mathrm{GS}^{+ +})$, where $W_\mathrm{GS}^{+ +}$ is a Hermitian matrix. Then, the expectation value of $S_{x,y}$ is calculated as
\begin{equation}
  \left(O^{++}(W_\mathrm{GS}^{++}),S_{x,y}O^{++}(W_\mathrm{GS}^{++})\right)= \sum_{\sigma,\tau=1}^N\Tr\left[W_\mathrm{GS}^{++} L^{+}_{xy,\sigma\tau} W_\mathrm{GS}^{++}L^{+}_{xy,\sigma\tau}\right].
\end{equation}
Here, $L^{+}_{xy,\sigma\tau}$ is a Hermitian matrix defined by \eqref{eq:def_of_L_pm}. By Lemma 11, $W_\mathrm{GS}^{++}$ is positive or negative definite. This, together with Lemma 8, implies that $\Tr\left[W_\mathrm{GS}^{++} L^{+}_{xy,\sigma\tau} W_\mathrm{GS}^{++}L^{+}_{xy,\sigma\tau}\right]>0$. Thus one finds that \newline ${\left(O^{++}(W_\mathrm{GS}^{++}),S_{x,y}O^{++}(W_\mathrm{GS}^{++})\right)>0}$. Similarly, if the ground state eigenoperator is in $\mathscr{O}^{- -}$ sector and expressed as $O^{--}(W_\mathrm{GS}^{--})$, then $\left(O^{--}(W_\mathrm{GS}^{--}),S_{x,y}O^{--}(W_\mathrm{GS}^{--})\right)>0$.
Therefore, we obtain (5) for any ground state.
\par
\bibliographystyle{apsrev4-1}
\end{document}